\documentclass[aps,prl,twocolumn,groupedaddress,nofootinbib,showpacs]{revtex4}
\usepackage{graphicx,epsf,color,amsmath,subfigure}
\newif\ifdraft
\drafttrue
\newif\ifpreprint
\preprinttrue

\def\fig#1{Fig.~{\ref{#1}}}
\def\Fig#1{Fig.~{\ref{#1}}}

\def\tab#1{Table~{\ref{#1}}}

\def\tabs#1#2{Tables~{\ref{#1}} and {\ref{#2}}}

\def\spa#1.#2{\left\langle#1\,#2\right\rangle}
\def\spb#1.#2{\left[#1\,#2\right]}

\def\eps{\epsilon}

\def\M{{\cal M}}
\def\K{{\cal K}}

\def\nt{{n_3}}
\def\ns{{n_0}}
\def\na{{n_2}}
\def\ntsq{{n_3^2}}
\def\nssq{{n_0^2}}
\def\nasq{{n_2^2}}

\def\eqn#1{Eq.~(\ref{#1})}

\def\NeqFour{{{\cal N}=4}}

\def\be{\begin{equation}}
\def\ee{\end{equation}}
\def\bea{\begin{eqnarray}}
\def\eea{\end{eqnarray}}
\def\ba{\begin{eqnarray}}
\def\ea{\end{eqnarray}}

\def\twoloop{{(2)}}

\newbox\charbox
\newbox\slabox
\def\s#1{{      
        \setbox\charbox=\hbox{$#1$}
        \setbox\slabox=\hbox{$/$}
        \dimen\charbox=\ht\slabox
        \advance\dimen\charbox by -\dp\slabox
        \advance\dimen\charbox by -\ht\charbox
        \advance\dimen\charbox by \dp\charbox
        \divide\dimen\charbox by 2
        \raise-\dimen\charbox\hbox to \wd\charbox{\hss/\hss}
        \llap{$#1$} }}

\def\LEH{{\cal L}_{\rm EH}}
\def\LGB{{\cal L}_{\rm GB}}
\def\LRHH{{\cal L}_{RH\!H}}
\def\LRRR{{\cal L}_{R^3}}

\begin{document}

\ifpreprint
\hbox{\hskip0.3cm UCLA/15/TEP/101 \hskip1.5cm SLAC--PUB--16332 \hskip1.5cm
CALT-2015-036 \hskip1.5cm \hfill CERN-PH-TH-2015-162}
\fi

\vskip0.3cm

\title{Evanescent Effects Can Alter Ultraviolet Divergences\\
in Quantum Gravity without Physical Consequences}

\author{Zvi~Bern${}^{abc}$, Clifford Cheung${}^b$, Huan-Hang Chi${}^d$, Scott~Davies${}^a$, 
Lance Dixon${}^{bd}$, Josh Nohle${}^a$}
\affiliation{
${}^a$Department of Physics and Astronomy,
   University of California at Los Angeles, Los Angeles, CA 90095, USA \\
${}^b$Walter Burke Institute for Theoretical Physics, California Institute of Technology, Pasadena, CA 91125, USA \\
${}^c$Department of Physics, CERN Theory Division, CH-1211 Geneva 23, Switzerland\\
${}^d$SLAC National Accelerator Laboratory, Stanford University,  Stanford, CA 94309, USA\\
}

\begin{abstract}
Evanescent operators such as the Gauss-Bonnet term have
vanishing perturbative matrix elements in exactly $D=4$ dimensions.
Similarly, evanescent fields do not propagate in $D=4$; a three-form field
is in this class, since it is dual to a cosmological-constant
contribution.  In this Letter, we show that evanescent operators and
fields modify the leading ultraviolet divergence in pure gravity.
To analyze the divergence, we compute the two-loop identical-helicity
four-graviton amplitude and determine the coefficient of the
associated (non-evanescent) $R^3$ counterterm studied long ago by
Goroff and Sagnotti.  We compare two pairs of theories that are dual in $D=4$:
gravity coupled to nothing or to three-form matter, and gravity
coupled to zero-form or to two-form matter.  Duff and van
Nieuwenhuizen showed that, curiously,
the one-loop conformal anomaly --- the
coefficient of the Gauss-Bonnet operator --- changes under $p$-form
duality transformations.  We concur, and also find that the
leading $R^3$ divergence changes under duality transformations.
Nevertheless, in both cases
the physical renormalized two-loop identical-helicity four-graviton
amplitude can be chosen to respect duality.  In particular, its
renormalization-scale dependence is unaltered. 
\end{abstract}

\pacs{04.65.+e, 11.15.Bt, 11.25.Db, 12.60.Jv \hspace{1cm}}

\maketitle

Although theories of quantum gravity have been studied for many
decades, basic questions about their ultraviolet (UV) structure persist.
One subtlety is the conformal
anomaly\footnote{Einstein gravity is not conformally invariant, so
  this is not an anomaly in the traditional sense.}, also known as the
Weyl or trace anomaly~\cite{ConformalAnomaly}.  At one loop, the
conformal anomaly provides the coefficient of the Gauss-Bonnet (GB) term.
The physical significance of this relationship has not been settled,
however.  In particular, Duff and van
Nieuwenhuizen showed that the conformal anomaly changes under duality
transformations of $p$-form fields, suggesting that theories related
through such transformations are quantum-mechanically
inequivalent~\cite{DuffInequivalence}.  In response, Siegel
argued that this effect is a gauge artifact and therefore not
physical~\cite{SiegelEquivalence}; Fradkin and Tseytlin
and Grisaru et al.\ have also argued that duality should hold at the quantum
level~\cite{FradkinTseytlinEquivalence}.  Furthermore, for $D=4$ external
states, one-loop divergences
in gravity theories coupled to two-form antisymmetric tensors are unchanged
under a duality transformation relating two-forms
to zero-form scalars~\cite{OneLoopEquivalence}.  However, as
we shall see, intuition based on one-loop analyses can be deceptive.

As established in the seminal work of 't~Hooft and
Veltman~\cite{HooftVeltman}, pure gravity is finite at one loop
because the only available counterterm is the GB term, which
integrates to zero in a topologically trivial background.  While
amplitudes with external matter fields diverge at one loop, amplitudes
with only external gravitons remain finite.  At two loops, however,
pure gravity diverges, as demonstrated explicitly by Goroff and
Sagnotti~\cite{GoroffSagnotti} and confirmed by van de
Ven~\cite{vandeVen}.

In this Letter, we investigate the UV properties of the two-loop
amplitude for scattering of four identical-helicity gravitons,
including the effect of $p$-form duality transformations.
We use dimensional regularization, which forces us to consider the
effects of evanescent operators like the GB term, which are legitimate
operators in $D$ dimensions but vanish (or are total derivatives) in
four dimensions.  We show that the GB counterterm is required
to cancel subdivergences and reproduce the 
two-loop counterterm coefficient found
previously~\cite{GoroffSagnotti,vandeVen}.

Evanescent operators are well-studied in gauge theory
(see e.g.~Ref.~\cite{Evanescent}), where they can modify
subleading corrections.  In contrast, we find that 
evanescent effects can alter the leading UV 
divergence in gravity.\footnote{Effects of the GB term have also been
studied in renormalizable, but non-unitary, $R^2$ gravity~\cite{R2GB}.}
Despite this change in the UV divergence, the physical dependence of the
renormalized amplitude on the renormalization scale remains unchanged.  
This break in the link between the UV divergence and the renormalization-scale
dependence is unlike familiar one-loop examples.  We arrive at a
similar conclusion when comparing the divergences and renormalization-scale
dependences in gravity coupled to scalars versus antisymmetric-tensor fields.

Pure gravity is defined by the Einstein-Hilbert Lagrangian,
\begin{align}
\LEH & = - \frac{2}{\kappa^2} \sqrt{-g} R\,,
\label{Lagrangian}
\end{align}
where $\kappa^2 = 32 \pi G_N = 32 \pi/M_P^2$ and the metric 
signature is $(+---)$.  We also augment
$\LEH$ by matter Lagrangians for one of the following:
$\ns$ scalars, $\na$ two-form fields (antisymmetric tensors) 
or $\nt$ three-form fields:
\begin{align}
{\cal L}_0 & =   \frac{1}{2}   \sqrt{-g}
    \sum_{j=1}^{\ns}
       \partial_\mu \phi_j \partial^\mu \phi_j\,, \nonumber  \\
{\cal L}_2 & =  \frac{1}{6} \sqrt{-g} 
 \sum_{j=1}^{\na}
       H_{j\,\mu\nu\rho} H_j^{\mu\nu\rho} \,,\nonumber  \\
{\cal L}_3 & = - \frac{1}{8} \sqrt{-g} \sum_{j=1}^{\nt}
      H_{j\, \mu\nu\rho\sigma} H_j^{\mu\nu\rho\sigma} \,.
\label{LagrangiansMatter}
\end{align}
Here $\phi_j$ is a scalar field and $H_{j\,\mu\nu\rho}$ and
$H_{j\,\mu\nu\rho\sigma}$ are the field-strengths of the two- and
three-form antisymmetric-tensor fields $A_{j\,\mu\nu}$ and $A_{j\,\mu\nu\rho}$.
The index $j$
labels distinct fields.  Standard gauge-fixing for the two- and three-form
actions, as well as for $\LEH$, leads to a nontrivial ghost structure.
We avoid such complications by using the generalized
unitarity method~\cite{UnitarityMethod,DDimensionalUnitarity,AllPlusQCD}, which
directly imposes appropriate $D$-dimensional physical-state projectors
on the on-shell states crossing unitarity cuts.

Under a duality transformation, in four dimensions the two-form field is equivalent to a
scalar:
\begin{equation}
H_{j\,\mu\nu\rho}\ \leftrightarrow\ \frac{i}{\sqrt{2}} \,
\varepsilon_{\mu\nu\rho\alpha} \, \partial^{\alpha}\phi_j \,,
\end{equation}
and the three-form field is equivalent to a cosmological-constant
contribution via
\begin{align}
H_{j\,\mu\nu\rho\sigma}\ \leftrightarrow\
\frac{2}{\sqrt{3}} \, \varepsilon_{\mu\nu\rho\sigma} \, 
 \frac{\sqrt{\Lambda_j}}{\kappa} \,.
\end{align}
As usual, we expand the graviton field around a flat-space background: 
$g_{\mu\nu} = \eta_{\mu\nu} + \kappa h_{\mu\nu}$.  Similarly, we expand the
scalar, two-form field and three-form field around trivial
background values.
It is interesting to note that the three-form field has
been proposed as a means for neutralizing the cosmological 
constant~\cite{Cosmology}. 

For a theory with $\ns$ scalars, $\na$ two-forms and $\nt$ three-forms 
coupled to gravity, the one-loop UV divergence takes the form
of the GB term~\cite{ConformalAnomaly,GoroffSagnotti,DuffInequivalence},
\begin{align}
\LGB\ =&\ \frac{1}{(4 \pi)^2} \frac{1}{\eps}
\Bigl( \frac{53}{90}+ \frac{\ns}{360} + \frac{91 \na}{360}
 - \frac{\nt}{2} \Bigr)\nonumber \\
& \times \sqrt{-g} (R^2 - 4 R_{\mu\nu}^2 + R_{\mu\nu\rho\sigma}^2) \,,
\label{GaussBonnetDivergence}
\end{align}
which is proportional to the conformal anomaly.  The
calculations of the conformal anomaly and of the UV
divergence are essentially the same, except that we replace a
graviton polarization tensor with a trace over indices.
Contracting \eqn{GaussBonnetDivergence} with four on-shell $D=4$
graviton polarization tensors gives zero. This is because the
GB combination is evanescent in $D=4$: It is a total derivative and vanishes
when integrated over a topologically trivial space; hence
pure Einstein gravity is finite at one loop~\cite{HooftVeltman}.
In a topologically nontrivial space, the integral
over the GB term gives the Euler characteristic.
When matter is added to the theory, the four-graviton
amplitude is still UV finite at one loop,
although divergences appear in
amplitudes with external matter states.

Using the unitarity method, we verified
\eqn{GaussBonnetDivergence} by considering the
one-loop four-graviton amplitude with external states in arbitrary
dimensions and internal ones in $D=4-2\eps$ dimensions.
On-shell scattering amplitudes are sensitive only to the coefficient of the
$R_{\mu\nu\rho\sigma}^2$ operator, because the $R^2$ and
$R^2_{\mu\nu}$ operators can be eliminated by field redefinitions
at leading order in the derivative expansion.  The GB combination is
especially simple to work with in dimensional regularization since
there are no propagator corrections in any dimension~\cite{Zwiebach}.

For the case of antisymmetric tensors coupled to gravity, another
relevant one-loop four-point divergence is that of two gravitons and
two antisymmetric tensors, generated by the operator,
\begin{equation}
\LRHH = \left(\frac{\kappa}{2}\right)^2{1 \over (4\pi)^2} \frac{1}{\eps}
\sqrt{-g} \sum_{j=1}^{\nt} R^{\mu\nu}_{\ \ \rho\sigma} H_{j\,\mu\nu\alpha}
H_j^{\alpha\rho\sigma} \,.
\label{RHHDivergence}
\end{equation}
Like the GB term, this operator is evanescent. In particular, in $D=4$,
we can dualize the antisymmetric tensors to scalars, which collapses the
Riemann tensor into the Ricci scalar and tensor.  Under field redefinitions,
they can be eliminated in favor of the dualized scalars, removing the one-loop
divergence in two-graviton two-antisymmetric-tensor amplitudes with
$D=4$ external states.  The four-scalar amplitude does diverge.

The change in \eqn{GaussBonnetDivergence} under duality transformations
is central to the claim by Duff and van Nieuwenhuizen of quantum
inequivalence under such transformations~\cite{DuffInequivalence}.
Here we analyze their effects on the two-loop amplitude.
First let us note that our
unitarity-based evaluation of \eqn{GaussBonnetDivergence} sews
together physical, gauge-invariant tree amplitudes.  This explicitly
demonstrates that the numerical coefficient of the
$R_{\mu\nu\rho\sigma}^2$ term in \eqn{GaussBonnetDivergence} is gauge
invariant, in contrast to implications of
Ref.~\cite{SiegelEquivalence}.  This gauge invariance suggests that
by two loops, \eqn{GaussBonnetDivergence} could lead to
duality-violating contributions to non-evanescent operators.  To see
if this happens, we must account for subdivergences and
renormalization.

At two loops, pure gravity diverges in $D=4$.  The coefficient of this
divergence was
determined by Goroff and Sagnotti~\cite{GoroffSagnotti} from a
three-point computation in the standard $\overline{\rm MS}$
regularization scheme and later confirmed by van de
Ven~\cite{vandeVen}:
\begin{equation}
\LRRR = 
-\frac{209}{1440} \left(\frac{\kappa}{2}\right)^2 \frac{1}{(4 \pi)^4} \frac{1}{\eps} 
 \sqrt{-g}\, R^{\alpha \beta}{\!}_{\gamma\delta} 
R^{\gamma \delta}{\!}_{\rho\sigma} R^{\rho\sigma}{\!}_{\alpha\beta} \,,
\label{GSDiv}
\end{equation}
where we account for the fact that
Refs.~\cite{GoroffSagnotti,vandeVen} define $\eps = 4 - D$ instead of
our $\eps = (4-D)/2$.  The divergence in \eqn{GSDiv} uses
four-dimensional identities to simplify it.

In order to reproduce the Goroff and Sagnotti result, we evaluate the
identical-helicity four-graviton amplitude. This is the simplest
amplitude containing the two-loop divergence (\ref{GSDiv}).  While a
four-point amplitude may seem to be unnecessarily complicated with
respect to a three-point function, there are several advantages to
considering an amplitude for a physical process with real momenta.
The first is that we can use the unitarity method to obtain a compact
integrand~\cite{UnitarityMethod}.  This method is particularly
efficient for identical-helicity particles, having been used to obtain
compact integrands for the gauge-theory case~\cite{AllPlusQCD}.  More
importantly, the question of quantum equivalence under duality
transformations can only be properly answered in the context of
physical observables, such as renormalized and infrared-subtracted $2
\rightarrow 2$ scattering amplitudes entering physical cross-sections.

To facilitate comparisons to the two-loop four-point amplitude,
we need the $R^3$ divergence (\ref{GSDiv}) inserted into 
the four-plus-helicity tree amplitude:
\begin{equation}
A_{R^3} =  \frac{209}{24} \frac{\K}{\eps}  \,,
\label{TwoLoopDivergenceGS}
\end{equation}
where 
\begin{equation}
\K \equiv \Bigl(\frac{\kappa}{2} \Bigr)^6 \frac{i}{(4 \pi)^4}
s t u \biggl(\frac{[1 2] [3 4]}
{\langle 1 2 \rangle \langle 3 4 \rangle} \biggr)^2 \,,
\label{KDef}
\end{equation}
and $s = (k_1 + k_2)^2$, $t = (k_2 + k_3)^2$ and $u = (k_1 + k_3)^2$ 
are the usual Mandelstam invariants.  The last factor is
a pure phase constructed from the spinor products $\langle a b \rangle$ 
and $[a b]$ defined in, for example, Ref.~\cite{Review}. 

\begin{figure}[tb]
\begin{center}
\subfigure[]{\includegraphics[scale=.28]{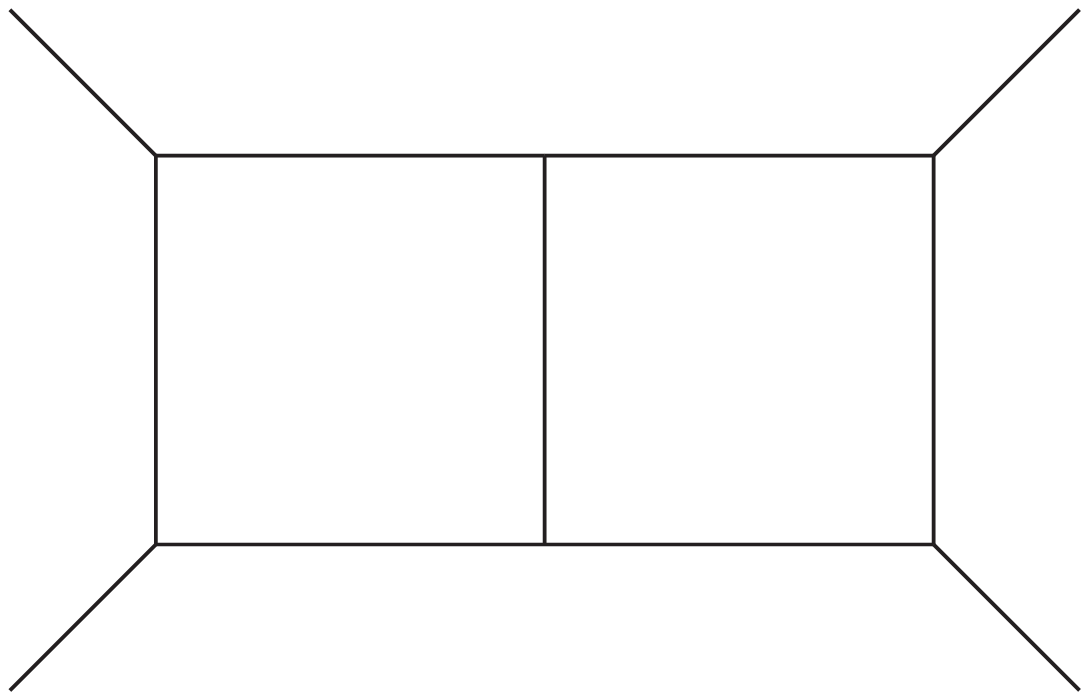}}
\hspace{.3cm}
\subfigure[]{\includegraphics[scale=.28]{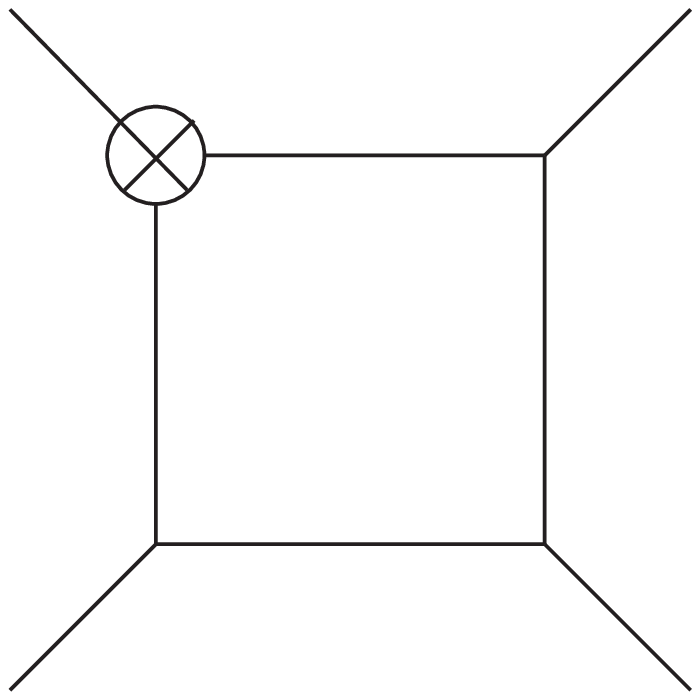}}
\hspace{.3cm}
\subfigure[]{\raisebox{.55cm}{\includegraphics[scale=.28]{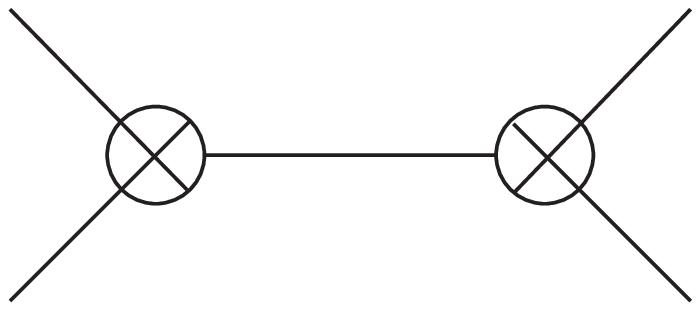}}}
\end{center}
\vskip -.7cm 
\caption[a]{\small Representative diagrams of the (a) bare,
(b) single-counterterm and (c) double-counterterm insertions.
\label{CounterSingleDoubleFigure}
}
\end{figure}

\begin{table}[t]
\vskip .4 cm
\def\hs{\hskip .4 cm }
\begin{tabular}{l c  c}
 &$1/\eps$ &  $\ln \mu^2$   \\[.1cm]
\hline
bare$\vphantom{\bigg|}$&$-\frac{3431}{5400}-\frac{199\nt}{30}+6\ntsq\hs$
&$-\frac{3431}{2700}-\frac{199\nt}{15}+12\ntsq$
\\[.15cm]
GB& $\frac{4\cdot 53-180\nt}{360} \cdot  \frac{2\cdot(13+180\nt)}{15}\hs$
&$\frac{689}{675}+\frac{199\nt}{15}-12\ntsq$
\\[.15cm]
GB$^2$&$24 \bigl(\frac{4\cdot 53 -180 \nt}{360}\bigr)^2\hs$
       & 0
           \\[.15cm]
total\hs
& $\frac{209}{24} - \frac{15}{2} \nt  \hs$
& $-\frac{1}{4}$ \\
\end{tabular}
\caption{Coefficients of the $1/\eps$ UV pole and of $\ln \mu^2$ in
  the identical-helicity four-graviton two-loop amplitude for pure
  gravity coupled to $\nt$ three forms.  We omit the overall factor of $\K$
  defined in \eqn{KDef}.  The first row gives the bare
  two-loop contribution, the second row the single GB-counterterm
  insertion at one loop, and the third row that of a double
  GB insertion at tree level.  The final row gives the total.
\label{DivergenceTableThreeForm}
}
\end{table}

\begin{table}[t]
\vskip .4 cm
\def\hs{\hskip .4 cm }
\begin{tabular}{l c  c}
 &  $1/\eps$ &  $\ln \mu^2$   \\[.1cm]
\hline 
bare
$\vphantom{\bigg|}$
& $ -\frac{3431}{5400} - \frac{277 \ns}{10800} + \frac{\nssq}{5400}\hs $
& $-\frac{3431}{2700} - \frac{277 \ns}{5400} + \frac{\nssq}{2700}$
\\[.15cm]
GB
    & $ \frac{4\cdot 53+ \ns}{360} \cdot  \frac{2\cdot(13-\ns)}{15}  \hs$
    & $\frac{689}{675} - \frac{199 \ns}{2700} - \frac{\nssq}{2700} $
\\[.15cm]
GB$^2$
       & $24 \bigl(\frac{4\cdot 53 + \ns}{360}\bigr)^2 $
       & 0             
           \\[.15cm]
total\hs
& $\frac{209}{24} - \frac{1}{48} \ns \hs$
& $-\frac{2+\ns}{8}$ \\
\end{tabular}
\caption{Coefficients of the $1/\eps$ UV pole and of $\ln \mu^2$ in
  the four-graviton amplitude for gravity coupled to $\ns$ scalars.
  The table follows the same format as \tab{DivergenceTableThreeForm}.
\label{DivergenceTableScalar}
}
\end{table}

\begin{table}[t]
\vskip .4 cm
\def\hs{\hskip .3 cm }
\begin{tabular}{l c  c}
$\null$ &  $1/\eps$ &  $\ln \mu^2$ \\ [.1cm]
\hline 
bare
$\vphantom{\bigg|}$ & $\hskip -0.15 cm 
   -\frac{3431}{5400} + \frac{8543 \na}{10800} + \frac{8281 \nasq}{5400}\hs$
& $ -\frac{3431}{2700} + \frac{8543 \na}{5400}  + \frac{8281 \nasq}{2700} $
\\[.15cm]
GB     
    & $ \frac{4\cdot 53+ 91 \na}{360} \cdot  \frac{2\cdot(13- 91 \na)}{15}\hs$
    & $\frac{689}{675} - \frac{18109 \na}{2700} - \frac{8281 \nasq}{2700} $
\\[.15cm]
GB$^2$                     
       & $24 \bigl(\frac{4\cdot 53 + 91 \na}{360}\bigr)^2 \hs $
       & 0             
           \\[.15cm]
$RH\!H$ 
       & $5 \na \hs$
       & $5 \na$
           \\[.15cm]
total\hs
& $\frac{209}{24} + \frac{299}{48} \na   \hs$
& $-\frac{2+\na}{8}$ \\
\end{tabular}
\caption{Coefficients of the $1/\eps$ UV pole and of $\ln \mu^2$ in
  the two-loop four-graviton amplitude for gravity coupled to $\na$
  antisymmetric-tensor fields. The table follows the same format as
  \tab{DivergenceTableThreeForm}.  The second-to-last row gives the
contribution of the $RH\!H$ counterterm inserted into the one-loop amplitude.
\label{DivergenceTableTwoForm}
}
\end{table}

\Fig{CounterSingleDoubleFigure} shows that there are three
types of contributions to consider: (a) the bare two-loop
contribution, (b) the one-loop single-counterterm subtraction and (c)
the double-counterterm subtraction. One might expect the net
subdivergence subtractions (b) and (c) each to be zero because there
are no corresponding $D=4$ one-loop divergences.  However, this is not
correct.  A careful analysis of the two-loop
integrands~\cite{LongPaper} reveals subdivergences associated with
the GB term (\ref{GaussBonnetDivergence}).  For the case of two-forms,
a subdivergence corresponding to $\LRHH$ in
\eqn{RHHDivergence} must also be subtracted.  In principle, when
three-forms are present, there might have been subdivergences
due to operators containing three-forms, but these do not
appear.  It is somewhat surprising that there are subdivergences at
two loops without any corresponding one-loop divergences in $D=4$.
However, because some legs external to the subdivergence are in $D$
dimensions, the cancellations that are specific to $D=4$ do not occur.

While Goroff and Sagnotti also subtracted subdivergences, they did so
integral by integral, rather than tracking the operator origin of the
subdivergences as we do.  Here we use dimensional regularization for
both infrared and UV divergences; we subtract the well-known infrared
singularities~\cite{IRPapers} from the final result.

We evaluate the bare and single-subtraction contributions via the
unitarity method.  We take the external legs to be identical-helicity
gravitons and each internal leg to be $D$-dimensional.  The bare integrand
obtained in this way is similar to integrands found earlier for
gauge theory~\cite{DDimensionalUnitarity,AllPlusQCD} and for
the ``double-copy'' theory containing a graviton, an antisymmetric tensor and a
dilaton~\cite{DoubleCopyGravity}. A key property of these integrands
is that they vanish when the loop momenta are taken to reside in
$D=4$, yet the amplitudes are still nonvanishing.  This phenomenon
is related to the observation by Bardeen and
Cangemi~\cite{AllPlusAnomaly} that the nonvanishing of
identical-helicity amplitudes is connected to an anomaly in the
self-dual sector.

We follow the same regularization prescriptions used in
Ref.~\cite{AllPlusQCD}, where algebraic manipulations on the integrand
are performed with $\eps < 0$.  We use the 't~Hooft-Veltman variant:
We place the external momenta and polarizations in $D=4$ and
take the loop momenta and internal states to reside in $D=4-2\eps$
dimensions. Here we focus on the UV divergences and defer
presentation of the integrands and finite terms
in the amplitudes to Ref.~\cite{LongPaper}.

We integrate over the loop momenta with the
same techniques used to obtain two-loop four-point
helicity amplitudes in QCD, including their finite
parts~\cite{TwoLoopQCDIntegration}.  As a cross check, we also directly
extract the UV divergences using masses to regulate the
infrared~\cite{DoubleCopyGravity}.

Consider first the case of $\nt$ three-forms coupled to gravity. In
\tab{DivergenceTableThreeForm}, we give both the divergence and
renormalization-scale dependence of each of the three components
illustrated in \fig{CounterSingleDoubleFigure}.  In the bare and
one-loop single-insertion components, the $\ln\mu^2$ dependence,
where $\mu$ is the renormalization scale, is proportional to the
UV divergence.  For the bare two-loop part, the $\ln\mu^2$
coefficient is twice the coefficient of the $1/\eps$ divergence.  For
the single counterterm, it is equal to the $1/\eps$ coefficient, and for
the double-insertion tree contribution, it vanishes.  This follows from
dimensional analysis of the loop integrals, with measure $\int d^{4-2\eps}\ell$
per loop, requiring an overall factor of $\mu^{2L\eps}$ at $L$ loops.
The counterterm subtractions are pure poles that do not carry such factors.
In the sum over terms, there is no simple relation between the
$1/\eps$ and the $\ln\mu^2$ coefficients, in contrast to many textbook
examples at one loop.

As seen from the last line of \tab{DivergenceTableThreeForm}, with no
three-form fields we match exactly the Goroff and Sagnotti
divergence~(\ref{TwoLoopDivergenceGS}).  The addition of $\nt$
three-form fields shifts the divergence from the pure gravity result.
One might think that this shift would lead to a physical change
in the scattering amplitudes through a different dependence on $\mu$.
However, the $\ln\mu^2$ column
of \tab{DivergenceTableThreeForm} shows that the $\nt$-dependence
of the bare and single-counterterm contributions precisely
cancels in the sum.  The scale dependence is therefore unaffected by three-form
fields.  The differences in the divergent parts can be removed
by adjusting the coefficient of the $1/\eps$ $R^3$ counterterm.
We have also obtained the amplitude's finite parts~\cite{LongPaper}.
Their form allows for a finite $R^3$ subtraction that completely eliminates
the effects of three-form fields in the two-loop renormalized
identical-helicity amplitude.

We now turn to the case of duality transformations between
antisymmetric-tensor fields and scalars.  In
\tabs{DivergenceTableScalar}{DivergenceTableTwoForm}, the coefficients
of $1/\eps$ and $\ln\mu^2$ terms are collected.  The tables show that,
while the individual components are quite different and the final
$1/\eps$ divergence changes under duality transformations, scalars and
two-forms have exactly the same renormalization-scale dependence.  As
for the case of three-forms, we find that the UV divergence does
depend on the field representations, but the renormalization-scale
dependence does not.  Again, finite subtractions can be found to make
the dual pair of renormalized amplitudes identical~\cite{LongPaper}.

From
Tables~\ref{DivergenceTableThreeForm}--\ref{DivergenceTableTwoForm},
we find that in all cases, the scale dependence in the
identical-helicity four-graviton amplitude follows a simple behavior:
\begin{equation}
\M^{\twoloop}_4 \Bigr|_{\ln\mu^2} =  - \K \,\frac{N_b-N_f}{8}\, \ln\mu^2\,,
\label{NbNf}
\end{equation}
where $N_b$ ($N_f$) is the number of bosonic (fermionic) four-dimensional
states in the theory.  We only computed \eqn{NbNf} explicitly for $N_f=0$, but
the identical-helicity graviton amplitude vanishes in supersymmetric
theories, forcing \eqn{NbNf} to be proportional to $N_b-N_f$.

The $\ln\mu^2$ dependence is clearly a more appropriate quantity for
deciding whether a theory should be thought of as nonrenormalizable.
If the coefficient of the $\ln\mu^2$ is nonvanishing, as is the case
for pure gravity, the coefficient will run, and we consider such a
theory to be nonrenormalizable.  Our result shows that instead of
focusing on the divergences, one should study the $\ln\mu^2$
coefficient to see if there is a principle that can be applied to set
it to zero.  One obvious useful principle is that renormalization
schemes should be chosen that maintain the equality of theories
related by duality transformations.

In this light, one might wonder if the recently-computed four-loop
divergence of pure $\NeqFour$ supergravity~\cite{N4FourLoop} is an
artifact of the particular $SU(4)$ formulation of the theory that was
used.  However, with the uniform mass infrared regulator used in that
calculation, extensive checks reveal that all subdivergences
cancel. Therefore the coefficient of $\ln\mu^2$ is proportional to
that of the $1/\eps$ divergence.  When matter multiplets are added
there are one-loop subdivergences, but those are not evanescent.  In
other formulations, it is possible that the divergences will change,
but we do not expect the $\ln\mu^2$ coefficients to change.

In summary, our investigation of the ultraviolet divergences of
nonsupersymmetric gravity reveals a number of striking features.  The
first is the nontrivial role of the conformal anomaly and the
associated evanescent Gauss-Bonnet term entering subdivergences.  It is
remarkable that a term that vanishes in four dimensions can
contribute directly to the leading divergence of a graviton amplitude.
Another important feature is that the integrand of the identical-helicity
amplitude vanishes if the loop momenta are taken to be four-dimensional;
this feature of identical-helicity amplitudes, which follows straightforwardly
from unitarity, is also tied to anomalous
behavior~\cite{AllPlusAnomaly}.  Similar connections to anomalous
behavior~\cite{MarcusAnomaly} were noted in the four-loop divergence
of $\NeqFour$ pure supergravity~\cite{N4FourLoop}.

A key lesson is that under duality transformations, the values of
two-loop divergences can change, contrary to the situation at one
loop~\cite{OneLoopEquivalence}.  However, the difference in these
divergences are unphysical, in the sense that they can be absorbed
into a redefinition of the coefficient of a local operator.  In other
words, our results for scattering amplitudes are consistent with
quantum equivalence under duality transformations when that
equivalence allows for the adjustment of coefficients of higher-dimension
operators.  The dependence on the renormalization scale does
not change under duality transformations in the examples we studied;
it is a more appropriate measure of the UV properties of the
theory. It would be quite interesting to establish this property
beyond two loops.
Together with recent examples of ultraviolet finiteness in
supergravity amplitudes, despite the existence of seemingly valid
counterterms~\cite{N4gravThreeLoops,N5FourLoop}, the results
summarized in this Letter show that much more remains to be learned
about both duality at the quantum level and the ultraviolet structure
of gravity theories.

\vskip .2 cm

We especially thank David Kosower, Radu Roiban, Augusto Sagnotti and
Raman Sundrum for many useful and interesting discussions and
suggestions.  We also thank Luis Alvarez-Gaume, John Joseph Carrasco,
Stanley Deser, Paolo Di Vecchia, Gary Horowitz, Henrik Johansson, Tim
Jones, Kelly Stelle and Mark Wise for helpful discussions.  This
material is based upon work supported by the Department of Energy
under Award Number DE-{S}C0009937 and contract DE--AC02--76SF00515 and
the Gordon and Betty Moore Foundation through Grant No.~776 to the
Caltech Moore Center for Theoretical Cosmology and Physics.  CC is
supported by a DOE Early Career Award DE-{S}C0010255 and a Sloan
Research Fellowship.  ZB and LD are grateful to the Simons Foundation
for support and to the Walter Burke Institute at Caltech for
hospitality.  CC and LD thank the Aspen Center for Physics and the NSF
Grant \#1066293 for hospitality.  SD and JN gratefully acknowledge
Mani Bhaumik for his generous support.


\end{document}